\documentclass[utf8]{frontiersSCNS} 

\setcitestyle{round} 
\usepackage{url,lineno,microtype,subcaption}
\usepackage[onehalfspacing]{setspace}

\def\keyFont{\fontsize{8}{11}\helveticabold }
\def\firstAuthorLast{Bisogni {et~al.}} 
\def\Authors{Susanna Bisogni\,$^{1,2,*}$, Alessandro Marconi\,$^{3,2}$, Guido Risaliti\,$^{3,2}$ and Elisabeta Lusso\,$^{4}$}
\def\Address{$^{1}$Harvard-Smithsonian Center for Astrophysics, Cambridge, MA, USA \\
$^{2}$INAF, Osservatorio Astrofisico di Arcetri, Firenze, Italy\\
$^{3}$Universit\`a degli Studi di Firenze, Dipartimento di Fisica e Astronomia, Sesto Fiorentino, Firenze, Italy \\
$^{4}$Centre for Extragalactic Astronomy, Department of Physics, Durham University, Durham, UK  }
\def\corrAuthor{Susanna Bisogni}

\def\corrEmail{susanna.bisogni@cfa.harvard.edu}

\begin{document}
\onecolumn
\firstpage{1}

\title[The effects of orientation on quasars spectra]{EW[OIII] as an orientation indicator for quasars: implications for the torus} 

\author[\firstAuthorLast ]{\Authors} 
\address{} 
\correspondance{} 

\extraAuth{}

\maketitle

\begin{abstract}

\section{}

We present an analysis of the average spectral properties of ~12,000 SDSS quasars as a function of accretion disc inclination, as measured from the equivalent width of the [O III] 5007\AA\ line. The use of this indicator on a large sample of quasars from the SDSS DR7 has proven the presence of orientation effects on the features of UV/optical spectra, confirming the presence of outflows in the NLR gas and that the geometry of the BLR is disc-like. Relying on the goodness of this indicator, we are now using it to investigate other bands/components of AGN. Specifically, the study of the UV/optical/IR SED of the same sample provides information on the obscuring “torus”. The SED shows a decrease of the IR fraction moving from face-on to edge-on sources, in agreement with models where the torus is co-axial with the accretion disc. Moreover, the fact we are able to observe the broad emission lines also in sources in an edge-on position, suggests that the torus is rather clumpy than smooth as in the Unified Model. The behaviour of the SED as a function of EW[OIII] is in agreement with the predictions of the clumpy torus models as well.

\tiny
 \keyFont{ \section{Keywords:} galaxies: active, galaxies: nuclei, galaxies: Seyfert, quasars: emission lines, quasars: general} 
\end{abstract}

\section{Introduction}

The fact we are not able to spatially resolve the inner regions of Active Galactic Nuclei (AGN), combined with their axisymmetric geometry \citep{AntonucciMiller1985, Antonucci1993}, can make it difficult to interpret their emissions.
The Unified Model predicts the orientation to be one of the main drivers of the diversification in quasars spectra. For this reason, an indicator of the inclination of the source with respect the line of sight of the observer is essential to get further in studying these objects.

Despite several quasars properties have been found to provide information on the inclination of the inner nucleus \citep{WillsBrowne1986,WillsBrotherton1995,Boroson2011,Decarli2011,VanGorkom2015}, we still lack an univocal measurement of this quantity. This problem is even harder when dealing with not-jetted objects, the most among quasars \citep[$>90\%$,][]{Padovani2011}, for which we can not rely on the presence of the strongly collimated radio-jets, directed perpendicularly to the accretion disc.

In order to give a more accurate description of the components surrounding the central engine and to understand where are the boundaries between one and another, we need to know which components are being intercepted by our line of sight.

Assuming that some of these inner components are characterised by a spherical geometry can often simplify the scenario, while at the same time misleading us. We use the emission lines coming from the Broad Line Region (BLR) to give an estimate of the mass of the central Super Massive Black Hole (SMBH), but in doing that we do not take properly into consideration the geometry of the BLR, i.e we use an average \emph{virial factor} $f$ to account for the uknown in the geometry and kinematics of the emitting region and we overlook the effects of orientation on the emission lines \citep{JarvisMcLure2006,Shen2013}. These measurements can then be used in turn to examine the relations between the SMBH and their host galaxies - one of the few tools available to understand the connection between structures on such different spatial scales - their uncertainties affecting these studies \citep[e.g. ][]{ShenKelly2010}.
If the BLR is characterised by a non-spherical geometry we are sistematically underestimating the BH masses in all the sources in which the velocity we intercept, i.e. the line width we measure, is only a fraction of the intrinsic velocity of the emitting gas orbiting around the SMBH.
The inclination of the source with respect to the line of sight is therefore crucial to both the understanding of how the nuclei work and how they affect the formation and evolution of galaxies in the Universe.
In this proceeding we show recent results on the optical spectra and we present a preliminary result on the Spectral Energy Distribution of quasars, that we obtained using the EW[OIII] as an orientation indicator.

\section{Orientation effects on emission features}

\subsection{Optical spectra}
Based on the properties of the [OIII] $5007$\AA~ line - negligibly contaminated by non-AGN processes and coming from the Narrow Line Region (NLR), whose dimensions ensure the isotropy of the emissions \citep{Mulchaey1994, Heckman2004} - and on the strong anisotropy of the continuum emitted by the optically-thick/geometrically-thin accretion disc \citep{ShakuraSunyaev1973}, we proposed the equivalent width (EW) of the [OIII] line, the ratio between the two luminosities, as an indicator of quasars orientation \citep{Risaliti2011,Bisogni2017}. 

In \cite{Risaliti2011} we examined the distribution of the observed EW[OIII] in a large sample of quasars from the SDSS DR5 ($\sim 6000$) and verified the presence of an orientation effect: the distribution shows a power law tail at the high EW[OIII] values that can not be ascribable to the intrinsic differences in the NLR among different objects, i.e. the intrinsic EW[OIII] distribution, the one we would observe if all the sources were seen in a face-on position.
The observed EW[OIII] distribution is a convolution of the intrinsic properties of the NLR emissions in the different objects, such as the ionising continuum and the covering factor of the clouds, and the effects due to their inclination angle. 

\begin{figure}
\begin{center}
\includegraphics[width=18cm]{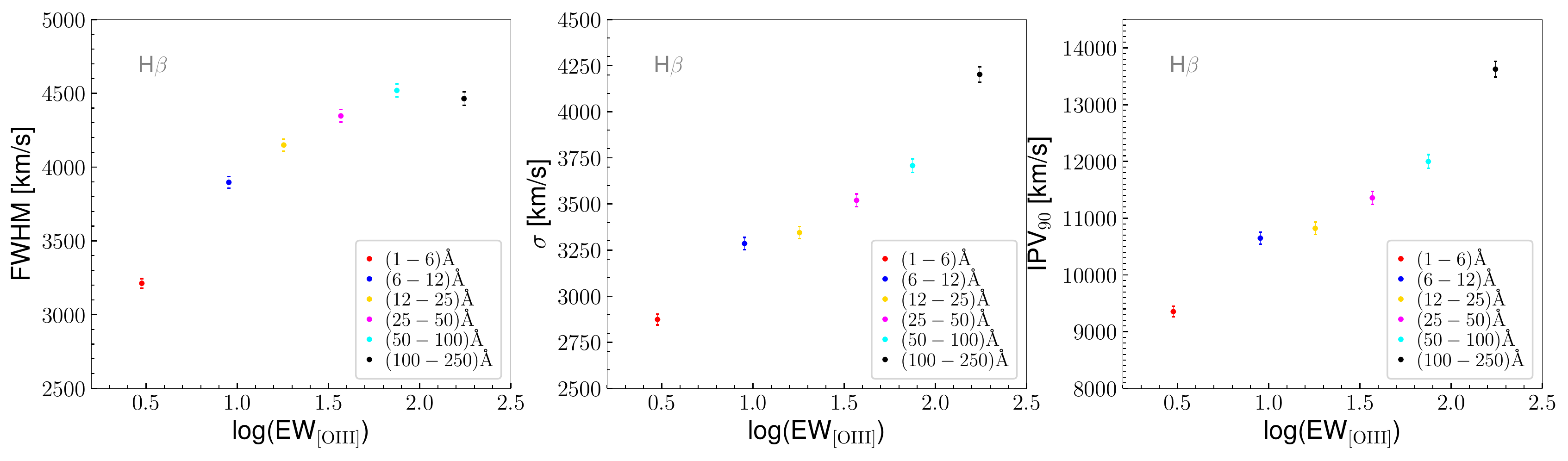}
\end{center}
\caption{Full Width Half Maximum, Inter-Percentile Velocity width and dispersion $\sigma$ as a function of the EW[OIII] for the H$\beta$ line. All these quantities, describing the rotational velocity of the gas orbiting around the central SMBH, increase moving from low to high EW[OIII] as expected if the BLR is disc-shaped and we are moving from face-on to edge-on positions.}\label{fig1}
\end{figure}

In \cite{Bisogni2017} we selected a larger sample of objects from the SDSS DR7 ($\sim 12000$), this time with the aim of looking for evidences of orientation effects in the optical spectra.
We split our sample in six bins of EW[OIII], each one corresponding to an inclination angle range. Within each bin, the spectra were stacked in order to produce a master spectrum.
We then analysed both the broad and the narrow emission lines as a function of EW[OIII], i.e of the inclination angle, finding orientation effects on both of them.
Fig. \ref{fig1} shows the presence of orientation effects on the broad component of H$\beta$: the width of the broad line, either represented by the line dispersion $\sigma$, the Full Width Half Maximum (FWHM) or the Inter-Percentile Velocity width (IPV), increases steadily when we move from low to high EW[OIII]. We found the same result for the other broad lines examined (H$\alpha$ and MgII, see \cite{Bisogni2017} for more details). This behaviour is what is expected if the BLR geometry is disc-like and we are moving from sources in a face-on position to sources in an edge-on position.


\begin{figure}
\begin{center}
\includegraphics[width=18cm]{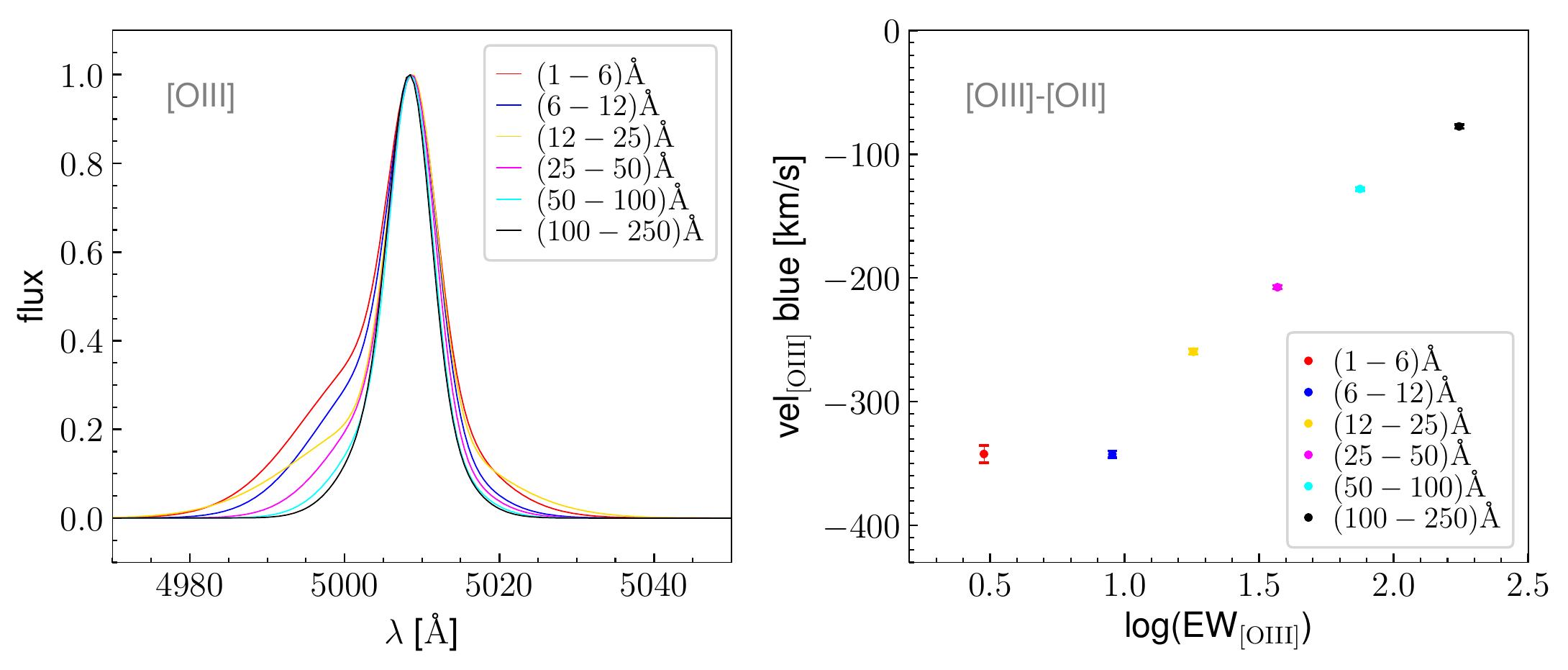}
\end{center}
\caption{Left panel: [OIII] $5007$\AA~ line profile as a function of EW[OIII]. The blue component of the line decreases moving from low to high EW[OIII]. Right panel: velocity shift of the blue component of the [OIII] line with respect to the velocity of the [OII] $\lambda 3727$ line, accounting for the systemic velocity of the host galaxy. The shift is decreasing (in modulus) when moving from low to high EW[OIII].  Both the trends are expected if we are moving from face-on to edge-on positions, where the outflow velocity component of the gas is not intercepted anymore by the line of sight.}\label{fig2}
\end{figure}

As for the narrow emission lines, we examined the [OIII] $\lambda 5007$\AA, the most prominent among them in the optical spectrum. This line is known to be contaminated by emissions coming from non-virialized gas, i.e. not orbiting around the central SMBH, but outflowing perpendicularly to the accretion disc \citep{Heckman1984, Boroson2005}. 
If the EW[OIII] is a good indicator of the inclination of the source, we should see the blue component of the line, emitted by outflowing gas, decreasing both in intensity and in velocity shift with respect to the nominal wavelength of the emission moving to high EW[OIII] values. Going from face-on to edge-on position in fact we are not intercepting anymore the outflow perpendicular to the accretion disc. This behaviour is found in the [OIII] line profile (Fig. \ref{fig2}).


We want to stress that within each EW[OIII] bin, therefore within each inclination angle range, the population of quasars is characterised by different SMBH masses, luminosities and accretion rates. These properties are considered among the main drivers of the variance in quasars spectra \citep{Marziani2003,Zamanov2002,ShenHo2014}. In our study however, as it is designed, even if the effects produced by these properties are present, they are diluited in the stacked representative spectra. 
As a confirmation that the orientation, even if not the only driver, plays a major role in the variance of quasars spectra, we found a clear trend of the Eigenvector 1, i.e. the anticorrelation between the FeII and [OIII] emissions intensity that \cite{BorosonGreen1992} identify as the main responsible for quasars spectral variance, with the EW[OIII].
Specifically, when we move from low to high EW[OIII], i.e. from low to high inclination angles, we see the [OIII] intensity increasing, while FeII emission becomes less and less intense. This can be explained in terms of orientation: the BLR shares the same anisotropy of the accretion disc and therefore the intensity of its emissions, in this case FeII, is decreased by a factor $\cos \theta$, i.e. decreases when moving to edge-on sources. On the other hand the [OIII] line appears as more evident in edge on positions because the luminosity of the continuum emitted by the accretion disc is decreased by the factor $\cos \theta$.

\subsection{Infrared emissions}

The observed EW[OIII] distribution has implications for the obscuring component as well.

The torus is depicted in the Unified Model as a smooth and toroidal structure that can reach $\sim 1- 10$ pc in size \citep{Burtscher2013}. If this is true then, there is a maximum inclination angle beyond which we are not able anymore to intercept the emissions coming from the very inner components, such as the continuum emitted by the accretion disc and the broad lines emitted by the BLR.
In this case, however, the observed EW[OIII] distribution would drop when the line of sight is starting to intercept the torus. This is not what we observe: the power law keeps going very steadily to the highest EW[OIII] values.
Moreover, we are intercepting broad emission lines in positions corresponding to high inclination angles.
Both these facts are not compatible with the torus being a smooth structure and rather suggest a clumpy structure.

To test the indicator and exploit its potential, we are now interested in investigating the infrared emissions.
 
We then collect photometric data for the same sample in the UV, optical and IR band to study the Spectral Energy Distributions (SED) of the sources.

\section{Sample and data analysis}
For the sample of $\sim 12000$ objects we selected from the SDSS DR7 the following photometric data are available:

\begin{itemize}
\item[-] Far Ultra Violet and Near Ultra Violet bands from \emph{Galaxy Evolution Explorer} (\emph{GALEX}) DR5 \citep{Bianchi2011}. 
\item[-] \emph{ugriz} SDSS photometric data from \cite{Shen2011}.
\item[-] The J, H and K bands from the \emph{Two Micron All-Sky Survey} (2MASS) \citep{Skrutskie2006}.
\item[-] The $3.4$, $4.6$, $12$ and $22\,\mu$m photometric data from the \emph{Wide-field Infrared urvey Explorer} (\emph{WISE}) \citep{Wright2010}.
\end{itemize}

We first correct all the magnitudes for Galactic extinction using the maps from the \cite{Schlegel1998}. Then, for each EW[OIII] bin, we use the same approach as for the optical spectroscopic data: we rest-frame the data according to the sources redshift and then we perform a stacking of the interpolated SED in order to produce a master SED, on which we can examine the effects produced by the orientation.
Before stacking them, we normalise each SED by dividing for the value of $\nu L_{\nu}$ at $\lambda=15 \, \mu$m, a reference wavelength in the mid-infrared spectral range, the band in which the torus emits. In doing that, we are normalizing the individual SED for the intrinsic differences of the torus and of other components in the different objects, such as the size and covering factor of the obscuring region, its distance from the central engine and the properties of the ionising continuum, whose emission is being reprocessed by the torus. This makes us able to compare the average behaviour of the obscuring structure at different inclinations with respect to the line of sight of the observer. The final SED are shown in Fig. \ref{fig3}.

\section{Results and discussion: implications for the obscuring torus}

\begin{figure}[h!]
\begin{minipage}[b]{.5\linewidth}
\centering\includegraphics[scale=0.6]{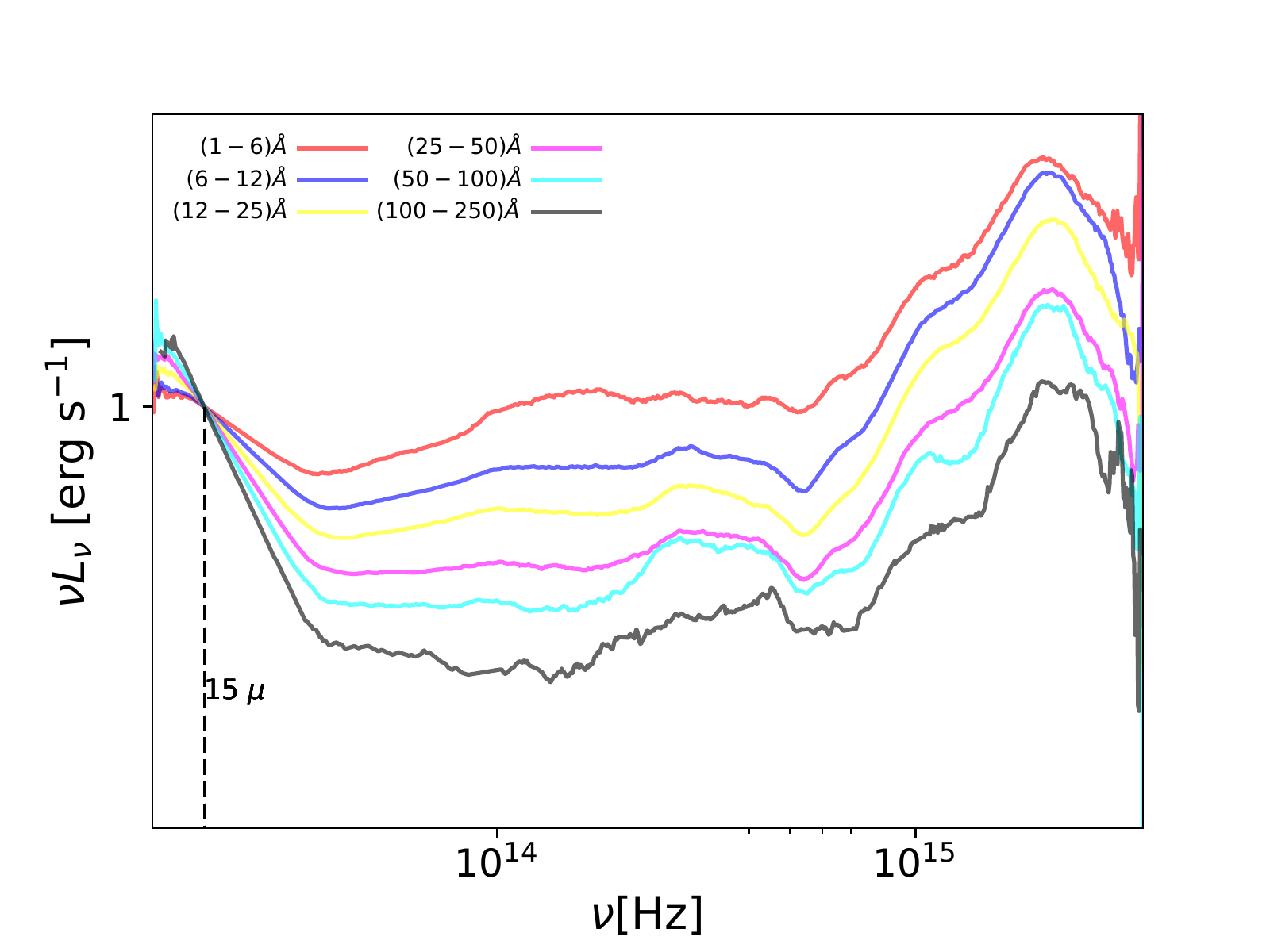}
\end{minipage}%
\begin{minipage}[b]{.5\linewidth}
\centering\includegraphics[scale=0.575]{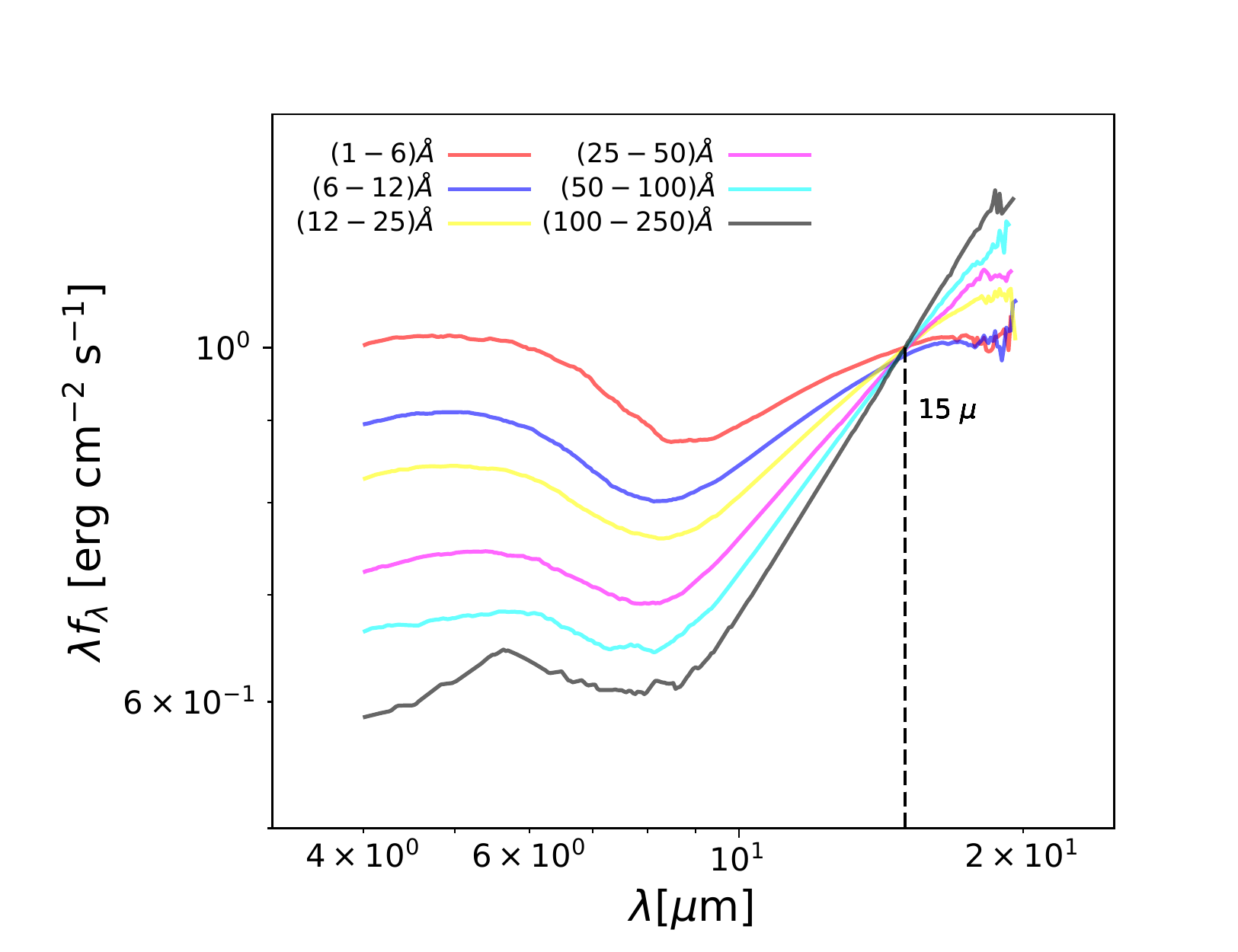}
\end{minipage}
\caption{(Left panel) Spectral Energy Distributions for the six EW[OIII] bins, corresponding to different inclination angle ranges. The master SED for a EW[OIII] bin was realized as follows: the photometric data for each source from the \emph{GALEX}, SDSS, 2MASS and \emph{WISE} surveys were corrected for Galactic reddening, rest-framed, interpolated on a common grid and normalized to the $\nu L_{\nu}$ value corresponding to the reference wavelength ($15\mu$m); for every channel in the grid we then selected the median value. (Right panel) Total flux for the six EW[OIII] bins in the spectral range in which the emission coming from the torus is predominant. SED corresponding to low EW[OIII] values are characterised by a shallower decrease in the emission at the shorter IR wavelengths with respect to the longer ones, while in the case of high EW[OIII] the decrease is steeper. This behaviour is in agreement with the clumpy torus models (see text for details).}\label{fig3}
\end{figure}

In the optical stacks corresponding to the highest inclination angles (high EW[OIII]) we are able to detect emissions from the BLR.
This evidence implies three possible scenarios: the absence of the torus, a torus that is mis-aligned with respect the plane of the accretion disc (and of the BLR) and a clumpy torus.
The first scenario is ruled out by the fact that the IR emission is clearly visible in the SED of the sample (Fig. \ref{fig3}).
As for the second one, we see the IR emission in the stacked SED decreasing progressively as a function of the indicator, defined through the anisotropic properties of the emission coming from the accretion disc itself. If torus and accretion disc were not co-axial, we would not see such an orderly behaviour.

The only scenario we are left with is therefore a clumpy torus, leading to a differention between type 1 and type 2 AGN due only to the photon escaping probability \citep{Elitzur2008}.
Due to the selection we performed (i.e. we selected blue objects, and verified that the continuum in our stacked optical spectra was not experiencing any reddening, see \cite{Bisogni2017}), when we are looking at sources with a high EW[OIII], i.e. with a high inclination angle, we are dealing with type 1 sources in which the BLR is intercepted through the dusty clouds of the torus.


We can compare our results with the clumpy models in literature \citep{Nenkova2008a, Nenkova2008b} that examine the infrared emission of the torus as a function of the inclination angle with respect to the observer.
If the torus is a clumpy structure, what we expect is that the IR emission at shorter wavelengths decreases progressively more than the ones at longer wavelengths when we are reaching edge-on position, due to the combination of an increasing number of clouds intercepted by the line of sight and of a higher absorption at the shorter than at the longer wavelengths \citep{Nenkova2008b}.

The behaviour of the stacked SED as a function of EW[OIII] confirms this scenario. At low EW[OIII] (low inclination angle) we are able to intercept the IR emissions coming from the inner clouds of the torus, that are directly illuminated by the ionising continuum, while at high EW[OIII] (high inclination angle) the IR emission coming from the inner clouds is shielded and we can detect it only after it is absorbed by the clouds in the outskirts of the torus. This produces the decrease in the flux at the shorter wavelengths, that becomes progressively more important for stacks corresponding to higher inclinations.

As a final verification that our results are not biased by any characteristics of the sample, we made the following checks:
1.) since our sample contains non-jetted as well as jetted quasars, the most extreme among them (blazars) could contaminate the part of the SED pertaining to the torus emission. We verified that the sub-sample composed by non-jetted quasars only gives the same result as the complete sample.\\
2.) $\sim 50\%$ of the objects in our sample has a redshift $z>0.47$. This is the critical value beyond which the normalisation flux at $15\,\mu$m is retrieved through an extrapolation rather than an interpolation of the SED. 
We verified that the analysis on the $z<0.47$ and $z>0.47$ sub-samples does not give different results. The only differences in the $z<0.47$ ($z>0.47$) sub-sample we recognise with respect to the complete sample SED are: a lower (higher) luminosity in Big Blue Bump (accretion disc), due to the fact that our sample is on average more luminous at higher redshifts, and a higher (lower) emission in the optical/NIR band, due to a higher contribution from the host galaxy for sources at lower redshift. We conclude that the extrapolation of the $15\,\mu$m flux in $z>0.47$ sources does not affect our results.

\section{Conclusions}

In this proceeding we summarise the results of the analysis on the optical spectra and we present the preliminary results of the analysis on the infrared emission of $12000$ sources of the SDSS DR7 as a function of the EW[OIII], a new orientation indicator.
We find that:
\begin{itemize}
\item[-] the BLR shares the same geometry of the accretion disc; we are intercepting the intrinsic velocity of the orbiting gas only when we are looking at sources in edge-on positions. If not properly taken into account, the orientation effects affecting the broad emissions lead to an underestimation of the SMBH virial masses in every position but the edge-on ones.
\item[-] the presence of outflowing gas in the NLR is clearly seen in the profile of the [OIII] $\lambda 5007$\AA~ as a function of the inclination angle. The blue component decreases both in intensity and in the shift with respect to the reference wavelength moving from face-on to edge-on positions.
\item[-] the preliminary analysis of the SED reveals a stronger decrease in the IR emission corresponding to the shorter wavelengths with respect to the longer ones when moving from low to high EW[OIII] values, as expected in the theoretical models for clumpy tori when moving from low to high inclination angles.
\end{itemize}

Further analysis is needed in order to investigate properly the emission coming from the torus. Starting from these first results, we are in the process of performing a SED fitting for each source in the sample with \emph{AGNfitter} \citep{CalistroRivera2016}. We will then be able to repeat the analysis on the representative SED, this time having also information on the single components contributing to the total emission.

We will also investigate the sources in our sample for which multiple observations are available (e.g. Stripe82, new BOSS spectra) in order to look for evidences of \emph{changing look} behaviours as a function of the EW[OIII]. If our interpretation of the data implying a clumpy structure for the torus is correct, we expect to see some of the sources that were included in our sample as Type 1 AGN changing to Type 2 objects at a different epoch. This behaviour is expected more frequently for sources with a high EW[OIII], where the orientation effect is dominant, but it is not excluded even for sources with low EW[OIII] values.

\section*{Conflict of Interest Statement}

The authors declare that the research was conducted in the absence of any commercial or financial relationships that could be construed as a potential conflict of interest.

\section*{Funding}
Support for this work was provided by the National Aeronautics and Space Administration through Chandra Award Number AR7-18013 X issued by the Chandra X-ray Observatory Center, which is operated by the Smithsonian Astrophysical Observatory for and on behalf of the National Aeronautics Space Administration under contract NAS8-03060. E.L. is supported by a European Union COFUND/Durham Junior Research Fellowship (under EU grant agreement no. 609412).

\bibliographystyle{frontiersinSCNS_ENG_HUMS} 
\bibliography{mybibOIII}

\end{document}